\documentclass[12pt]{article}
\usepackage{graphicx,epsf,amsmath,amsfonts,amssymb,amsbsy}

\def\i{\mbox{i}}
\def\\i{\mbox{\scriptsize{i}}}
\def\d{\mbox{d}}

\title{On the transverse size of hadrons at asymptotically high energies}
\author{M. L. Nekrasov\\
{\small\it Institute for High Energy Physics, NRC ``Kurchatov
Institute'',}  \vspace*{-4\baselineskip}\\
{\small\it Protvino 142281, Russia} }
\date{}
\begin{document}
\maketitle
\begin{abstract}
We show that Gribov diffusion of the partons in the impact parameter
plane, which leads to the square-root-of-logarithmic growth of the
transverse size of the hadrons, can occur only simultaneously with a
similar diffusion in the transverse-momentum space. At the same time, a
restriction of the partons in the transverse momenta entails an increase
in their propagation in the impact parameter plane. Ultimately this
leads to a logarithmic growth of the transverse size of hadrons at
asymptotically high energies.
\end{abstract}

\section{Introduction}\label{sec1}

Certain characteristics of interactions of hadrons at high energies are
directly related to their transverse sizes. These include the slope
parameter of elastic peripheral scattering and the total cross section
for strong hadron interactions. For the latter quantity there is a
restriction known as Froissart bound,
\begin{equation}\label{F1}
\sigma_t (s) \le \frac{4\pi}{t_0} \, \ln^2 \! s\,, \quad s \to \infty
\,,
\end{equation} 
derived \cite{Froissart,Martin1,Martin2} from the general
principles of local field theory, such as unitarity and analyticity
($t_0$ is the rightmost point on the Martin ellipse). The mode of
saturation of this bound corresponds to the physical picture of a black
disk with a logarithmically growing radius, $R\sim \ln s\,$ as $\,s
\!\to\! \infty$, see e.g. \cite{Gribov2}. However, the dynamic cause of
the bound (\ref{F1}) is unknown, and it is not known whether it is
saturated. In particular, there is currently no solution to these
problems in terms of QCD. In this regard, it is of interest to identify
independent dynamic conditions that lead to the aforementioned physical
picture.

Below we show that the logarithmic growth of the transverse size of the
hadron interaction region occurs as a consequence of the restriction of
the partons transverse momenta in the fast free-moving hadrons. (The
latter ones represent the hadrons in the limit case of soft collisions
at high energies.) The justification is very general in nature and is
based on the Heisenberg uncertainty relation. However, we use this
fundamental relation in a specific form that  corresponds to the problem
under consideration. Namely, we find that in systems developing by the
cascade decays, the product of variance of the coordinates of final
particles, counted from the geometric center, and the variance of
corresponding momenta is proportional to the number $N$ of the decays.
So, both variances can grow proportionally to $\sqrt{N}$ with increasing
$N$. However, if one of the variances is restricted for some dynamic
reasons, this entails an increase in the growth of the other variance,
so that their product remains proportional to $N$. 

In the case of fast free-moving hadrons, we assume, following
\cite{Feynman,Gribov1,Ioffe}, that the maximum number $N$ of the decays
in cascades increases with the energy as $\,\ln s$. During the decays,
partons lose longitudinal momenta and simultaneously propagate in the
transverse directions. We denote the transverse coordinate and momentum
of the last parton in the cascade as ${\cal R}$ and ${\cal K}$,
respectively. Then, as long as the propagation is not restricted, the
variances $\Delta {\cal R}$ and $\Delta {\cal K}$, actually the root
mean square (RMS) values of ${\cal R}$ and ${\cal K}$, both increase
with the energy as $\sqrt{\,\ln s}$. Hence, the transverse radius $R$ of
the hadron increases as $\sqrt{\,\ln s}$. However, if $\Delta {\cal K}$
becomes restricted, then due to the uncertainty relation $\Delta {\cal
R}$ must grow as $\,\ln s$. This means $R \sim \ln s\,$ as~$\,s \to
\infty$. 

In Sec.~\ref{sec2}, we give formal derivation of the relation $\Delta
{\cal R} \, \Delta {\cal K} \ge N$ in the systems developing by the
cascade decays. The appropriate scenarios of the evolution of the
partons are considered in Sec.~\ref{sec3}. In Sec.~\ref{sec4}, we
discuss
the~results.

\section{Uncertainty relation for cascade processes}\label{sec2}

First, we recall that the uncertainty relation resulting from the
commutation relation 
\begin{equation}\label{F2}
[\hat{x}_i,\hat{k}_j] = \i \delta_{ij}  \,,
\end{equation} 
takes the well-known form $\Delta x_i \, \Delta k_i \ge 1/2$ for the
one-dimensional motion only. In the case of arbitrary motion of the
particle in $n$-dimensional space, it has the form
\begin{equation}\label{F3}
\Delta x \, \Delta k \ge n/2\,,
\end{equation} 
where $(\Delta x)^2 = \langle \sum \hat x_i^2 \rangle - \sum \langle
\hat x_i \rangle^2$ and similarly for $\Delta k$. The actual value of
the product of variances in (\ref{F3}) depends on the state in which the
system is. In the case of vacuum-like state with Gaussian wave function,
the product is minimal.\footnote{Recall that in the Weyl approach, the
uncertainty relation is derived from (\ref{F2}) as a consequence of the
non-negativity of the norm of the vector $\hat{a}^{-}|\psi\!>$, where
$\hat{a}^{-}\! = (\hat{k} - \i \mu^2\hat{x})/\!\sqrt{2\mu^2}$. The
minimization of the uncertainty relation is achieved at the zero norm of
this vector, i.e.~in the case of a wave function satisfying the equation
$\hat{a}^{-}\psi = 0$.} The explicit form of the appropriate wave
function in the coordinate and momentum representations is as follows: 
\begin{eqnarray}\label{F4}
<\!\vec{x}\,|\psi\!> &\equiv& \psi (\vec{x}) =
\left(\mu^2/\pi\right)^{n/4} e^{\displaystyle -\vec{x}^{\,2}\mu^2/2}\,,
\\\label{F5}
<\!\vec{k}\,|\psi\!>  &\equiv& \widetilde\psi (\vec{k}) =
\left(4\pi/\mu^2\right)^{n/4} 
e^{\displaystyle -\vec{k}\,^{2}/(2\mu^2)}\,.
\end{eqnarray} 
Here we assume that the particle is on average at the origin and at
rest, $\mu^2$ is a~para\-meter of variance,
\begin{equation}\label{F6}
<\!\psi\,|\hat{\vec x}^{\,2}|\psi\!> = \frac{n}{2}\,\mu^{-2}\,, \qquad
<\!\psi\,|\hat{\vec {k}}\,^{2}|\psi\!> = \frac{n}{2}\,\mu^{2}\,.
\end{equation} 

Next, we consider a system of particles in two-dimensional space,
developing by sequential splittings. This system will simulate the
behavior of the partons in the perpendicular plane relative to the
direction of motion of the fast-moving hadron.\footnote{The possibility
of separate consideration of the perpendicular and longitudinal motion
of the partons is justified by the factorization of the longitudinal and
transverse variables in the scattering amplitudes at high energies in
multi-peripheral kinematics \cite{Gribov2}.} We assume that each
particle of the system emits a similar particle once, as it occurs in
the multi-peripheral comb \cite{Gribov1}. However, we will not consider
the system in the field theoretical approach and will confine ourselves
to non-relativistic quantum mechanical consideration, taking advantage
of the fact that in a fast-moving hadron real movements in the
perpendicular plane are frozen. An advantage of this approach is that
the use of the wave function initially allows us to track the
quantum-mechanical uncertainties. Correspondingly, we characterize each
particle $i$ by the radius vector $\vec {r}_i$ counted from the position
of the parent particle. Herewith, the radius vector of the first
particle ($i\!=\!1$) is counted from the geometric center of the system.
Assuming that all particles are independent and distributed relative to
the parent particles with a common variance, the wave function of the
system with $N$ splittings is 
\begin{equation}\label{F7}
\Psi_{N}(\{\vec{r}_i\}) = 
\prod_{i=1}^{N} \psi (\vec{r}_i) =
\left( \mu^2/\pi \right)^{N/2} \exp \left\{ 
- \frac{\mu^2}{2} \sum_{i=1}^{N} \vec{r}_i^{\,2} \right\} .
\end{equation} 
Notice the translational non-invariance of this modus of specifying the
system, which is a pay for accounting for ``quantum trembling'' of
the particles. We also note the factorization of the wave function and
emphasize that the use of sequentially relative coordinates provides 
separation of variables. When describing, for example, in terms of
absolute distances from the geometric center, the variables are mixed
due to scalar products of the vectors. Moreover, in the given
parameterization each particle is found effectively in its own
two-dimensional space, which causes the analogy of the system of $N$
particles with one particle in $2N$-dimensional space. 

In the momentum representation wave function (\ref{F7}) takes the form
\begin{equation}\label{F8}
\widetilde{\Psi}_{N}(\{\vec{k}_i\}) =
\prod_{i=1}^{N} \widetilde\psi (\vec{k}_i) =
\left( 4\pi/\mu^2 \right)^{N/2} \exp \left\{ 
- \frac{1}{2\mu^2} \sum_{i=1}^{N} \vec{k}_i^{\,2} \right\} .
\end{equation} 
Here $\vec{ k}_i$ are the local momenta of the particles relative to the
particles-parents. Their averages are zero, and all they are distributed
with variance $\mu^{2}$. In the context of the parton model $\mu^{2}$ is
determined by the scale of transverse momenta arising at the parton
splittings. From the viewpoint of our model $\mu^{2}$ is connected with
the effective mass $m$ of the particles, generated at each splitting.
Since the system is non-relativistic, we put $\mu^2 = m \omega$, where
$\omega$ is an auxiliary parameter such that $m \gg \mu$. One may
imagine occurrence of $\omega$ as the result of action of a weak
($\omega/\mu \ll 1$) oscillatory potential emanating from parent
particles. Such a potential does not change the Gaussian structure of
the wave function and provides non-relativistic properties of the
system. Both of these conditions are what we need from the system ($m$
and $\omega$ are not used further). 

Consider now the last particle formed at the $N$-th splitting (decay).
Its position relative to the geometric center is described by the vector
\begin{equation}\label{F9}
\vec{\cal R} = \sum_{i=1}^{N} \vec{r}_i \,.
\end{equation} 
Actually $\vec{\cal R}$ makes sense of a collective variable
characterizing evolution of the system as a whole. The average of
$\vec{\cal R}$ is zero (the system does not shift relative to the
geometric center). Therefore the variance of $\vec{\cal R}$ coincides
with the mean of the square:
\begin{equation}\label{F10}
(\Delta {\cal R})^2 = 
\langle \, {\vec{\cal R}} ^{2} \, \rangle =\;
<\!\Psi | \sum_{i=1}^{N} \hat{\vec{r}}_i^{\,2} +
\sum_{i \not= j}^{N} \hat{\vec{r}}_i \hat{\vec{r}}_j | \Psi > .
\end{equation} 
The first sum in (\ref{F10}) contains $N$ terms, the second one about
$N^2$. So $\Delta {\cal R}$ cannot grow faster than $N$. If the
particles are not correlated, the second sum vanishes and $\Delta {\cal
R} \sim \sqrt{N}$. With the wave function (\ref{F7}), we have
\begin{equation}\label{F11}
(\Delta {\cal R})^2 \,=\, N \mu^{-2}\,.
\end{equation} 

The collective momentum variable is a vector dual to (\ref{F9}), 
\begin{equation}\label{F12}
\vec{\cal K} = \sum_{i=1}^{N} \vec{k}_i \,.
\end{equation} 
Since the system is non-relativistic, $\vec{\cal K}$ has the meaning of
the momentum in the center-of-mass frame of the particle formed at the
$N$-th decay, provided that the effective masses of the particles are
equal each other. Since $\vec{\cal K}$ is distributed around zero (the
momentum is conserved), its variance is
\begin{equation}\label{F13}
(\Delta {\cal K})^2 = \langle \, {\vec{\cal K}} ^{2} \, \rangle =\;
<\!\Psi | \sum_{i=1}^{N} \hat{\vec{k}}_i^{\,2} +
\sum_{i \not= j}^{N} \hat{\vec{k}}_i \hat{\vec{k}}_j | \Psi > .
\end{equation} 
If the particles are not correlated, then $\Delta {\cal K} \sim
\sqrt{N}$,
and with the wave function (\ref{F8}) we have
\begin{equation}\label{F14}
(\Delta {\cal K})^2 \,=\, N \mu^{2}\,.
\end{equation} 
The growth of $\Delta {\cal K}$ with the growth of $N$ means an
accumulation of uncertainties of momenta during sequential splitting.
The $\Delta K$ characterizes the size of the system in the transverse
momentum space. 

From (\ref{F11}) and (\ref{F14}) we get
\begin{equation}\label{F15}
\Delta {\cal R} \, \Delta {\cal K} = N\,.
\end{equation} 
If the  wave functions were not Gaussian, we would get inequality in
(\ref{F15}), similar to (\ref{F3}). In the general case, this follows
from the observation that according to (\ref{F2}) the operators of
collective variables satisfy the commutation relation 
\begin{equation}\label{F16}
[\hat{\cal R}_\alpha,\hat{\cal K}_\beta] = 
\i N \delta_{\alpha\beta}  \,.
\end{equation} 
Here $\alpha$, $\beta$ are components of the vectors in the
two-dimensional space. Acting further according to the standard scheme
and taking into account the two-dimensionality of space, we come to 
\begin{equation}\label{F17}
\Delta {\cal R} \, \Delta {\cal K} \ge N\,.
\end{equation} 

This relation has far-reaching consequences. In particular, in the case
of a negative correlation in (\ref{F10}) or (\ref{F13}), which restrains
the growth of one of the variances, the growth of the other variance
should increase. So if $\Delta {\cal K} \sim 1$ as $N \to \infty$, then
(\ref{F17}) requires that there should be positive correlations in
(\ref{F10}) resulting in the behavior $\Delta {\cal R} \sim N$. Of
course, the wave function in this case is rearranged and cannot be
Gaussian. 

To characterize the behavior of the system as a whole, we introduce the
probability density distributions over collective variables. They will
describe the distributions of the last particle in the cascade decays.
In the configuration space, the probability density normalized to unity
is  
\begin{equation}\label{F18}
F_N(\vec{\cal R}) = 
\int \Bigl(\,\prod_i^N \d \vec{r}_i \Bigr) \,
\delta( \vec{\cal R} - \sum_i^N \vec{r}_i ) \; 
|\Psi_{N} (\{\vec{r}_i\})|^2\,.
\end{equation} 
In the case of Gaussian wave function (\ref{F7}), this gives
\begin{equation}\label{F19}
F_N(\vec{\cal R}) = \frac{\mu^2}{\pi N} \exp \left\{ 
- \frac{\vec{\cal R}^2 \mu^2}{N} \right\} ,
\end{equation} 
\begin{equation}\label{F20}
(\Delta {\cal R})^2 = \langle {\cal R}^2 \rangle = N \mu^{-2} \,.  
\end{equation} 
In the momentum space the probability density distribution is 
\begin{equation}\label{F21}
\!\!\!\!\!\!\!\!\!\!\!\!
{\cal F}_N(\vec {\cal K}) = 
\int \Bigl[\,\prod_i^N \frac{\d \vec{k}_i}{(2\pi)^2} \Bigr] \,
(2\pi)^2 \delta( \vec{\cal K} - \sum_i^N \vec{k}_i ) \; 
|\widetilde{\Psi}_{N}(\{\vec{k}_i\})|^2\,.
\end{equation} 
Substituting (\ref{F8}), we get
\begin{equation}\label{F22}
{\cal F}_N(\vec{\cal K}) = \frac{4\pi}{N\mu^2} \exp \left\{ 
- \frac{\vec{\cal K}^2}{N \mu^2} \right\},
\end{equation}
\begin{equation}\label{F23}
(\Delta {\cal K})^2 = \langle {\cal K}^2 \rangle = N \mu^{2}\,. 
\end{equation} 

Let us discuss the results. First of all, we emphasize that the Gaussian
form and the factor $N^{-1}$ under the exponent in both distributions,
(\ref{F19}) and (\ref{F22}), are the consequences of a free nature of
particle splittings, without restrictions and correlation. A remarkable
property of a system with this behavior is that its evolution with
increasing $N$ is similar to a diffusion process. This follows from the
observation that (\ref{F19}) and (\ref{F22}) are the fundamental
solutions to the two-dimensional diffusion equation in which $N$ plays
the r{\^{o}}le of diffusion time. Earlier, the analogy with diffusion
was used \cite{Gribov1} as the basis for determining the parton
distribution in the impact parameter plane, with the parton rapidity
playing the r{\^{o}}le of the time.\footnote{In fact both definitions of
the time are equivalent since the number $N$ of splittings is
proportional to the shift of the partons rapidity. Below we discuss this
point in more detail.} Simultaneously \cite{Gribov1} suggested that the
diffusion does not occur in the transverse momentum space. However, in
view of uncertainty relation (\ref{F17}), the latter suggestion is
incompatible with diffuse propagation in configuration space. The
diffusion can only occur simultaneously in both spaces. If there is no
diffusion expansion in the transverse momentum space and $\Delta {\cal
K}$ is independent of $N$, then by virtue of (\ref{F17}) $\Delta {\cal
R}$ should increase as $N$. This means that partons movement is
correlated. Such a scenario can be realized if ${\cal F}_N$ becomes
independent of $N$ at $N \to \infty$.

\section{Parton distributions in the transverse plane}\label{sec3}

Now we turn to the parton distributions in the real hadrons. Recall that
they are formed due to cascade decays, during which fast moving partons
give rise to slow ones. Simultaneously partons acquire local transverse
momenta relative to the parent partons. With respect to the center of
mass, these momenta are summed, as the system is non-relativistic in the
transverse directions. So, the distribution over transverse momenta
should be based on the summed local momenta. Unfortunately, this fact is
typically ignored, which distorts the expected distribution pattern.
Here we investigate this issue in the first approximation, based on the
parton model.  

We start with the remark that the number of partons is actually not
fixed in the fast-moving hadron. Therefore, the hadrons are
characterized by a set of the wave functions. Given this, the
probability distributions by collective variable $\vec{\cal R}$ is
written as 
\begin{equation}\label{F24} 
F(\vec{\cal R}) = \sum_{N=1}^{N_{max}}  |c_N|^2 
F_N(\vec{\cal R})\,,
\end{equation} 
and similarly for ${\cal F}(\vec{\cal K})$. Recall that $\vec{\cal R}$
and $\vec{\cal K}$ are the impact parameter and transverse momentum,
respectively, of the last partons in the cascades with $N$ decays, and
$N_{max}$ is a maximum number of the decays. If several cascades are
present simultaneously, we consider them united through appropriate
generalization of (\ref{F18}) and (\ref{F21}). We also neglect the
differences in the positions of the centers of different cascades and
between the initial transverse momenta. These simplifications are
inessential for our analysis since ultimately we are interested in
asymptotic properties of the distributions. 

The coefficients $|c_N|^2$ determine the probabilities that the hadron
consists of $N$ partons. They satisfy the relation
\begin{equation}\label{F25}
\sum_{N=1}^{N_{max}} |c_N|^2 = 1 \,.
\end{equation} 
A priori, $|c_N|^2$ are unknown. So $F$ and ${\cal F}$ are unknown, too,
whatever the partial distributions $F_N$ and ${\cal F}_N$ may be. We
only know that at large ${\cal R}$ and ${\cal K}$ the $F$ and ${\cal F}$
are well approximated by $F_{N_{max}}$ and ${\cal F}_{N_{max}}$, the
widest partial distributions. At the same time, the inner area is
dominated by narrower partial distributions.  

In general, the variances $\Delta {\cal R}$ and $\Delta {\cal K}$ are
the averages of the partial variances. So if $F_{N}$ and ${\cal F}_{N}$
are Gaussian, then we have by virtue of (\ref{F20}) and  (\ref{F24}) 
\begin{equation}\label{F26}
(\Delta {\cal R})^2 = \sum_{N=1}^{N_{max}} |c_N|^2 N \mu^{-2} =
\bar{N} \mu^{-2} \,.
\end{equation} 
Here $\bar{N}$ is the average number of the partons. Putting $\bar{N} =
\varkappa N_{max}$ with $\varkappa < 1$, we have
\begin{equation}\label{F27}
(\Delta {\cal R})^2 = \varkappa  \, \mu^{-2} N_{max} \,,
\end{equation} 
and smilarly
\begin{equation}\label{F28}
(\Delta {\cal K})^2 = \varkappa  \, \mu^{2} N_{max} \,.  
\end{equation} 

Further estimates may be obtained based on a relation between the number
$N$ of the decays  and the rapidity $\eta$ of the last parton in the
cascade. Following \cite{Gribov1,Ioffe}, we use $N = \gamma (\eta_P\!-\!
\eta)$, where $\gamma$ is a dimensionless factor and $\eta_P \approx \ln
2P/\mu$ is the maximal rapidity, $P$ is the hadron momentum, $\mu$ is a
dimensional scale. This relation permits to determine the distribution
of the partons by impact parameter $\rho$ depending on the rapidity of
the partons. Really, leaving in (\ref{F24}) the contributions of only
cascades that involve partons with rapidity $\eta$, and considering the
partial distributions truncated on these partons, we arrive at the
distribution 
\begin{equation}\label{F29}
\Phi_{\eta}(\rho) = 
\frac{C(\eta)\,\mu^2}{\pi \gamma (\eta_P\!-\! \eta)}
\exp \left\{ - \frac{\rho^2 \mu^2}{\gamma (\eta_P\!-\! \eta)} \right\}.
\end{equation} 
Here $\rho$ is a collective variable defined by (\ref{F9}) with the sum
up to $N_{\eta} = \gamma (\eta_P\!-\! \eta)$. The truncation in the
partial distributions implies that the internal summation in (\ref{F18})
is carried out up to $N_{\eta}$, where $N_{\eta} \leq N$. The result is
given by (\ref{F19}) with ${\cal R}$ replaced by $\rho$ and $N$ by
$N_{\eta}$. Formula (\ref{F29}) exactly reproduces distribution (12) of
\cite{Gribov1}, obtained there by analysing the front of the ``diffusion
wave''. It is obvious that partons with the minimal rapidity $\eta
\approx 0$ have the widest distribution, and such partons complete the
evolution in the cascades. In this limit case $\rho$ coincides with
${\cal R}$, and (\ref{F29}) becomes $F_{N_{max}}({\cal R})$ exclusive of
$C(0)$. In a similar way, we can derive the distribution by the
transverse momenta $k_{\perp}$ of the partons depending on their
rapidity. It looks identical to (\ref{F29}) with the replacement of
$\rho^2 \mu^2$ by $k_{\perp} / \mu^2$.  

From the above discussion, we have $N_{max} \! = \! \gamma \ln 2 P/\mu$,
and (\ref{F27}) gives $\Delta {\cal R}\! = \!\mu^{-1} \!
\sqrt{\varkappa\gamma \ln 2P/\mu}$. The latter estimate determines the
transverse size of the hadron, 
\begin{equation}\label{F30}
R \sim \sqrt{ \ln s} \,, \quad s \to \infty \,.
\end{equation}
It is worth noting that (\ref{F30}) reproduces the well-known result for
the radius of hadronic interactions caused by pomeron exchange
\cite{Gribov2}.

Another important corollary of (\ref{F27}) is the independence from the
energy of spatial density of the partons in the transverse projection.
Really, the hadron cross-sectional area is estimated as $S = \pi (\Delta
{\cal R})^2$. Hence, the two-dimensional density is $\varrho \equiv
\bar{N}/S = \mu^2/\pi$. We can roughly estimate $\varrho$ by relating
$\Delta {\cal R}$ to the slope of the diffraction cone of elastic hadron
scattering. Namely, we use well-known formula $B = \langle b^2
\rangle/2$, where $\langle b^2 \rangle$ is the average of the square of
the impact parameter, and following \cite{Petrov} we use estimate
$\langle b^2 \rangle = 2 (\Delta {\cal R})^2$.\footnote{The latter
relation was obtained as a condition that the valence cores of the
colliding hadrons cease to overlap in the impact parameter plane with
increasing the energy. The condition itself was considered as the reason
for the change in the growth mode of the slope $B$ from linear to
approximately logarithmic. We assume that the latter mode is actually
supported due to the similar relationship between $\langle b^2 \rangle$
and $(\Delta {\cal R})^2$.} In turn, in Regge theory $B = B_0 + 2
\alpha'(0) \ln s/s_0$. Gathering formulas, we get at $P \to \infty$ 
\begin{equation}\label{F31}
\varrho \;\approx\; \frac{\varkappa\gamma}{4\pi\alpha'(0)}\,. 
\end{equation} 
Assuming that partons carry on average half the momentum of the
partons-parents, we have $\gamma = (\ln 2)^{-1}$ \cite{Gribov2}. The
pomeron slope $\alpha'(0)$ is not exactly known. We use values
$\alpha'(0) = 0.165$ GeV$^{-2}$ and $\alpha'(0) = 0.321$ GeV$^{-2}$
obtained in different analysis schemes of various $pp$ and $p \bar{p}$
elastic scattering data including TOTEM, see \cite{DL1} and \cite{DL2},
respectively. This leads to $\varrho \approx \varkappa (\mbox{0.2--0.3
fm})^{-2}$. In both cases $\mu \approx \, \varkappa^{1/2} \,$GeV. Given
this, (\ref{F27}) and (\ref{F28}) are written as 
\begin{equation}\label{F32}
(\Delta {\cal R})^2 \approx (0.2 \, \mbox{fm})^2 \times N_{max}\,,
\end{equation} 
\begin{equation}\label{F33}
(\Delta {\cal K})^2 \approx 
\varkappa^{2} \, \mbox{GeV}^{2} \times N_{max}\,.
\end{equation} 

The above estimates seem reasonable. We emphasize that they are obtained
in the diffusive mode of the propagation of partons in the transverse
directions. Note the presence of the factor $\varkappa^{2}$ in
(\ref{F33}). With finite $N_{max}$, it permits to parametrically reduce
the estimate for $\Delta {\cal K}$ to almost any value below 1~GeV. It
should be also noted that the ``diffusion'' in the momentum space is
$\varkappa^{2}$ times slower than in the configuration space. 

An alternative mode of the behavior of partons occurs when their
propagation in the transverse directions is restricted in one of the
spaces, either spatial or momentum. There is a simple way to define the
proper space. Really, we have seen that in the diffusion mode the
spatial transverse density of the partons is independent from the
momentum of the hadron. Therefore, each parton occupies an area with a
constant average diameter regardless of the energy. This means that
hadrons can swell up unrestrictedly as new partons are formed with
increasing energy. On the contrary, an unrestricted growth of the
transverse momenta is incompatible with the hadron integrity. Hence, the
transverse momenta must be restricted. 

So, we assume that $\Delta {\cal K}$ cannot exceed some bound, say
$\Delta {\cal K}_{0}$. This implies that when $\Delta {\cal K}$
approaches $\Delta {\cal K}_{0}$, the wave function is radically
rearranged. This case was discussed in the end of Sec.~\ref{sec2}.
Given the uncertainty relation (\ref{F17}), this leads to the behavior 
\begin{equation}\label{F34}
\Delta {\cal R} \sim  (\Delta {\cal K}_{0})^{-1} \, N_{max} \,.
\end{equation}
In turn, this implies a logarithmic growth of the radius of the hadron
interactions, 
\begin{equation}\label{F35}
R \sim \ln s \,, \quad s \to \infty \,.
\end{equation}

In view of (\ref{F10}) and (\ref{F13}), the above behavior means
appearance of the constructive correlations in the configuration space
and destructive correlations in the momentum space, whereas (\ref{F32})
and (\ref{F33}) imply the uncorrelated motion. The transition to
(\ref{F34})
and $\Delta {\cal K} \sim \Delta {\cal K}_{0}$ implies a mode change in
which partons begin to ``feel'' the presence of each other. Accordingly,
the wave function of the system can no longer be presented in a
factorized form. The reason underlying this mode change is unclear. It
may be associated with the increase with the energy of the longitudinal
space occupied by the hadron \cite{Low,Petrov1,Petrov2} (effect of
uncertainty in its localization). As a result, the distances between the
partons are effectively increased, which means increased coupling. The
latter effect may be the cause of the correlation. However, further
research is needed to clarify this issue.

\section{Discusion}\label{sec4}

We have investigated parton distributions in the transverse directions
in the fast free-moving hadrons (in the limit case of soft collisions at
high energies). We have noted that from the view of the transverse
projection the patrons can be represented as quasi-particles
spontaneously multiplying through sequential splittings (cascade
decays). The effective masses of the quasi-particles are determined by
the scale of the transverse momenta arising from the splittings of
actual partons. Thus, the quasi-particles split without loss of mass.
The propagation of the quasi-particles occurs in the process of their
decays by means of including new local areas in the configuration and
momentum spaces, counting from the parent quasi-particles. The sizes of
the areas are controlled by the uncertainty relation. Ultimately, the
coordinates of the quasi-particles relative to the center of the system
are determined by summing local coordinates of all their parent
quasi-particles. Similarly, since the system is non-relativistic in the
transverse directions, the transverse momenta of the quasi-particles
relative to the center-of-mass are determined by summing the local
transverse momenta of the parent quasi-particles. The balance of the
propagation in the configuration and momentum spaces is controlled by
the uncertainty relation formulated in a special form relevant for
systems developing through the cascade decays. 

In general, the resulting transverse propagation of the quasi-particles,
in essence the partons, can be either uncorrelated or correlated. In the
former case their RMS transverse coordinates and momenta 
relative to the hadron center, both evolve by the diffusion law (however
somewhat slower in the momentum space). Namely, they increase as
$N_{max}^{1/2}$, where $N_{max}$ is the maximum number of the decays in
the cascades. So, if $N_{max} \sim \ln P$ \cite{Feynman,Gribov1,Ioffe},
where $P$ is the hadron momentum, then the RMS transverse coordinates
and momenta increase as $\sqrt{\ln P}$. However, if the transverse
momenta are restricted for some dynamic reason, then this means that a
correlation appears between the partons. In this case the RMS transverse
coordinates grow as $N_{max}$, i.e.~as $\ln P$. The latter behavior
means a logarithmic growth of the transverse size of the hadron and a
similar growth of the radius of strong hadron interactions. 

It is worth noting that the $\sqrt{\ln P}$ growth of the radius of
strong hadron interactions is actually long known as a consequence of
Regge behavior (with linear trajectories at small $t$) of the amplitude
of elastic hadron scattering \cite{Gribov2}. Gribov was the first to
explain this effect in the framework of the parton model as a
consequence of the diffusion of partons in the transverse projection
\cite{Gribov1}. But he did not take into account the diffusion of the
partons in the transverse momentum space. However, if the propagation of
the partons is restricted in the latter space, then this indicates the
appearance of a correlation between the partons. The correlation, in
turn, leads to an increase in the propagation of partons in the impact
parameter plane, and ultimately to the above-mentioned growth of the
transverse size of hadrons as $\ln P$. From the point of view of Regge
approach the latter effect can be interpreted as a transition to the
mode of multi-pomeron exchange.

We carried out our study in the framework of the parton model. This
implies that we did not take into account specific features of virtual
quarks and gluons. However, our outcomes are based on general physical
laws. Therefore, we expect them to be observed regardless of the dynamic
nature of the patrons, at least in a first approximation. Let us recall
that the dynamics of hadrons in the soft limit is currently outside the
real QCD capabilities. This is reflected, in particular, in the
inability of QCD to enforce the Froissart bound (at the current level of
research). So identifying the consequences of general principles remains
an urgent task in studying the behavior of the hadrons in the soft
limit.

We conclude that our results are of undoubted interest from the point of
view of QCD-modeling of the hadrons at high energies in the soft limit.
From the viewpoint of Froissart bound, our result about the logarithmic
growth of the transverse size of hadrons due to restriction of RMS
transverse momenta, means a detection of an independent dynamic
condition
for saturation of the bound.

\end{document}